\documentstyle[aps,epsfig,floats,12pt]{revtex}
\tightenlines
\begin{document}
\def\x{{\bf x}}
\def\k{{\bf k}}
\def\H{{\cal H}}
\def\ba{\begin{eqnarray}}
\def\ea{\end{eqnarray}}
\def\be{\begin{equation}}
\def\ee{\end{equation}}
\def\ob{{\Omega _{b,0}}}
\def\oc{{\Omega _{c,0}}}
\def\tr{{\rm tr}}
\font\boldsym=cmmib12\skewchar\boldsym='177
\textfont9=\boldsym \scriptfont9=\boldsym \scriptscriptfont9=\boldsym

\mathchardef\xib="0918
\title{Primordial Isocurvature Perturbations: Testing the Adiabaticity
of the CMB Anisotropy\footnote{To appear in the proceedings of the
CAPP2000 conference held at Verbier Switzerland, July 2000.
}}
\author{Martin Bucher, Kavilan Moodley and Neil Turok}
\address{DAMTP, Centre for Mathematical Sciences, University of Cambridge\\
Wilberforce Road, Cambridge CB3 0WA, United Kingdom}

\maketitle
\begin{abstract}
Prospects for testing the adiabaticity of the primordial 
cosmological perturbations using MAP and PLANCK are 
evaluated. The most general cosmological perturbation
in a universe with just baryons, photons, neutrinos,
and a cold dark matter (CDM) component is described. In addition
to the familiar adiabatic mode, there are four nonsingular
isocurvature modes: a baryon isocurvature mode, a CDM 
isocurvature mode, a neutrino density isocurvature mode, and
a neutrino velocity isocurvature mode. The most general
perturbation is described by a $5\times 5$, positive-definite, symmetric,
matrix-valued function of wave number whose off-diagonal elements
represent the correlations between the above mentioned modes. We
found that when three modes and their correlations are admitted,
the fractional uncertainties in the cosmological parameters
and amplitudes of the isocurvature modes and their correlations
become of order unity. These degeneracies, however, can be 
broken with the polarization information provided by PLANCK,
reducing the uncertainties to below the ten percent level.
Polarization is thus crucial to testing the adiabaticity 
of the primordial fluctuations. 
\end{abstract}

\vskip 14pt 

One of the fundamental challenges of the new cosmology is to establish
the underlying character of the primordial perturbations. Long before
inflation was proposed, it was pointed out that in their simplest form
the primordial perturbations would be adiabatic and Gaussian. It was
also pointed out that a scale-free power spectrum, the so-called
Harrison-Zeldovich-Peebles power spectrum \cite{HZP} , was an aesthetically
preferred form and also seemed to account for the then available data
quite well.

The subject of this contribution is to propose a test of the first of these two
hypotheses: that the primordial perturbations were adiabatic in
character. In this context adiabatic means that the primordial stress-energy of
the universe was governed by a single, spatially uniform equation of
state --- in other words, that on surfaces of constant temperature the
densities of the various components (e.g. baryons, CDM, neutrinos, etc.)
are uniform and that these components share a common velocity field. 

In order to test the hypothesis of adiabaticity, it is necessary to
contemplate models with non-adiabatic perturbations and then place
constraints on the possible amplitudes of these non-adiabatic
modes by comparing with the data. Most of the work on
how to interpret the future CMB data has assumed a simple one-field
inflationary model, for which the fluctuations are adiabatic absent
new physics, and a simple power law for the power spectrum
\cite{paramest}. A host of
cosmological parameters (such as $H_0, \Omega_\Lambda, \Omega_b,
\Omega_k, n_s$ and $\tau_{reion}$) 
are allowed to vary, and Bayesian statistical analysis is applied to
determine the values of these undetermined cosmological parameters and
their corresponding uncertainties. In many of the papers of this genre
that discuss how to analyse the future CMB data, the primary emphasis is placed
on how well one will be able to measure the various undetermined
cosmological parameters and very little discussion is devoted to
testing the underlying assumptions. Rather
it is assumed that should the real data not be described by a template
taken from the assumed class, this error would naturally become self-evident. 

Perhaps more interesting and of more profound significance to
discovering how the perturbations from homogeneity and isotropy were
first generated in the early universe is to determine the fundamental
character of the primordial perturbations. Some examples of questions
that one would wish to resolve include the following: Were the primordial
perturbations entirely adiabatic in character, or were isocurvature
modes excited as well? Were the primordial perturbations exactly
Gaussian? Were they `scalar,' or were `vector' and `tensor' modes
excited as well? Is an `acausal' theory such as inflation required, or
could the perturbations have been imprinted as a continual process?
Were `decaying' modes excited as well?

To answer these questions will require new approaches to 
analysing the forthcoming CMB data. The work reported in this
proceeding constitutes a modest step in this direction. In our approach
we make two simplifying assumptions: First, we assume a universe with
no new physics: a universe with only photons, baryons (and their associated
electrons), neutrinos, and a cold dark matter component. Second, we
assume that the perturbations were `primordial', meaning that they
were initially excited well before
recombination so that any decaying modes have had ample opportunity to
die away. The first hypothesis excludes, for example, theories of
structure formation based on field ordering such as cosmic strings
where the dynamics of some scalar or other order parameter field
evolves thus perturbing the other components \cite{pst,durrer}. 

Bucher {\it et al.} \cite{BMTa} 
analysed the most general cosmological perturbation
possible under these assumptions using synchronous gauge and found in
addition to the two gauge modes,  
five regular modes and two decaying modes for each
wavenumber {\bf k}: 
\begin{enumerate}
\item An adiabatic growing mode. 
\item An adiabatic decaying mode. 
\item A baryon isocurvature mode. 
\item A baryon velocity decaying mode. 
\item A CDM isocurvature mode. 
\item A neutrino density isocurvature mode. 
\item A neutrino velocity isocurvature mode. 
\end{enumerate} 

The leading order terms for all perturbation 
variables in the series expansion solution, which is valid at
early times and on superhorizon scales, is presented in Table I
for the regular modes listed above. 
Here $R_\gamma $ and $R_\nu $ represent the fractional contribution
of photons and neutrinos to the total density at early times
deep within the radiation dominated epoch, respectively, 
while $\ob$ and $\oc$ represent the present day fractional 
contribution to the total density of baryons and cold dark matter, respectively.
The first four modes have already been discussed extensively in the
literature. The observational consequences of the 
neutrino density and neutrino velocity modes, although
implicit in the work of Rebhan and
Schwarz\cite{nnewb} and of Challinor and Lasenby\cite{nnewc}, however,
have only recently been explored \cite{BMTa}. 

\def\ob{{\Omega _{b,0}}}
\def\oc{{\Omega _{c,0}}}
\def\R{ {R_\nu {R_\gamma}^{-1} }}
\def\RIA{(5+4R_\nu )^{-1}}
\def\RIB{(15+4R_\nu )^{-1} }
\begin{table}[!t]
\centering
\begin{tabular}{|| c | c | c | c | c | c ||}
\hline
\rule[-3mm]{0mm}{8mm}& AD & BI & CDMI & NID & NIV \cr
\hline\hline
\rule[-3mm]{0mm}{8mm} $h$ & 
$ {1\over 2}k^2\tau ^2  
$&$ 4\ob \tau$ 
&$4 \oc \tau
$&${1 \over 10}\ob \R k^2\tau ^3
$&${3\over 2}\ob {\R }k\tau ^2$ \cr 
\hline 
\rule[-3mm]{0mm}{8mm} 
$\eta $&$ 1 $&$-{2\over 3}\ob \tau $&$-{2\over 3}\oc \tau
$&$-{1 \over 6} R_\nu \RIB k^2\tau ^2
$&$-{4 \over 3} R_\nu \RIA k\tau$ \cr 
\hline 
\rule[-3mm]{0mm}{8mm}
$\delta_c $&$ -{1\over 4}k^2\tau ^2 $&$-2\ob \tau $&$1
$&$-{1\over 20}\ob \R k^2\tau ^3
$&$-{3 \over 4}\ob {\R}k\tau ^2$ \cr 
\hline 
\rule[-3mm]{0mm}{8mm}
$\delta_b $&$ -{1\over 4}k^2\tau ^2 $&$1 $&$ -2\oc \tau $&${1\over
8}{\R} k^2\tau ^2 
$&${\R}k\tau$ \cr
\hline 
\rule[-3mm]{0mm}{8mm}
$\delta_\gamma $&$ -{1\over 3}k^2\tau ^2 $&$-{8\over 3}\ob \tau $&$-{8\over
3}\oc \tau $&$ -{\R}
$&${4\over 3}{\R}k\tau$ \cr
\hline 
\rule[-3mm]{0mm}{8mm}
$\delta_\nu $&$ -{1\over 3}k^2\tau ^2 $&$-{8\over 3}\ob \tau $&$-{8\over
3}\oc \tau $&$1$&$-{4\over 3}k\tau$ \cr
\hline 
\rule[-3mm]{0mm}{8mm}
$\theta_c $&$ 0 $&$0  $&$0 $&$0 $&$0$ \cr
\hline 
\rule[-3mm]{0mm}{8mm}
$\theta_{\gamma b} $&$ -{1\over 36}k^4\tau ^3 $&$-{1\over 3}\ob k^2\tau ^2 
$&$-{1\over 3}\oc k^2\tau ^2 $&$ -{1\over 4}{\R}k^2\tau
$&$-{\R}k $ \cr
\hline 
\rule[-3mm]{0mm}{8mm}
$\theta_\nu $&$ -{1\over 36}\left[23+4R_\nu \over 15+4R_\nu \right]
k^4\tau ^3
$&$-{1\over 3}\ob k^2\tau ^2 $&$-{1\over 3}\oc k^2\tau ^2 $&${1\over 4}k^2\tau
$&$k$ \cr
%& & & & & \cr
\hline 
\end{tabular}
\vskip 12pt
\caption{For each mode the leading term in the power series expansion
in conformal time is shown, indicating the leading behavior on
superhorizon scales in the radiation era. For all these mode a
scale-invariant spectrum of large-angle CMB anisotropies corresponds 
to $P(k) \sim k^{-3}$.}
\end{table}

We next provide a brief qualitative discussion of these modes. For
more details the reader is referred to \cite{BMTa}. 
For the adiabatic mode, as previously mentioned, all components
obey a common equation of state and share a common velocity field. The
ratios of baryons to photons, CDM to photons, and neutrinos to photons
do not vary spatially. For the adiabatic `decaying' mode the perturbation
becomes singular as $t \to 0$. In the CDM \cite{bond} and baryon 
\cite{pjepa,bondb,cen} isocurvature modes, the ratios of CDM to
photons and of baryons to photons,
respectively, vary spatially. Initially, far into the radiation
dominated era, these components, idealized to be non-relativistic
particles whose density scales as $\rho \sim a^{-3}$ (as opposed to
$\rho \sim a^{-4}$ for radiation) contribute negligibly to the total
stress-energy. However, as the universe becomes matter dominated,
these variations translate into variations in the equation of state
that generate metric perturbations which in turn generate further
perturbations in the densities and velocities of all the components. 
For the baryon velocity `decaying' mode, the velocity of the baryons
relative to the photons is rapidly damped by the Thomson scattering of
photons off the baryons.

For the neutrino isocurvature modes the neutrino sector is perturbed
relative to the radiation sector (i.e., the photons and other
components that are strongly coupled to them at early times through
Thomson scattering). The perturbations are initially of equal
magnitude but in opposite directions, so that the total energy density
and momentum density vanishes.

Let us first discuss the neutrino density mode. If neutrinos and
photons evolved identically, no metric perturbation would
subsequently result. But as a mode enters the horizon, the neutrinos
free stream while the photons behave as a perfect fluid because of
Thomson scattering off electrons. This differential behavior leads to
perturbations in the total stress-energy, which generate metric
perturbations, which in turn generate perturbations in all 
components. The case of the neutrino velocity mode is
similar. Initially, the momentum densities cancel. However, because of
the differential dynamics upon horizon crossing, stress-energy
perturbations arise, which then source metric perturbations, which in turn
generate perturbations in the other components. One may have suspected
that this mode would be singular because the `velocity' mode in the
adiabatic sector is decaying and thus singular. But because of the
cancellation initially this too is a nonsingular mode. 

In Figure 1 the CMB anisotropies predicted for the various modes are
indicated. The CDM and baryon isocurvature modes predict CMB spectra
of the same shape, to within a fraction of a percent, so only
the baryon isocurvature mode is shown and studied below. 
Of the isocurvature modes, the
baryon isocurvature and CDM isocurvature modes have greatly
suppressed power on small scales relative to large scales 
while the neutrino density
isocurvature mode exhibits a rise at $\ell \approx 75$ leading to a plateau
before the first Doppler peak. However, interestingly, the neutrino
velocity isocurvature mode lacks these features, rather having the same
qualitative behavior as the adiabatic growing mode. 

\begin{figure} 
\centerline{\epsfig{file=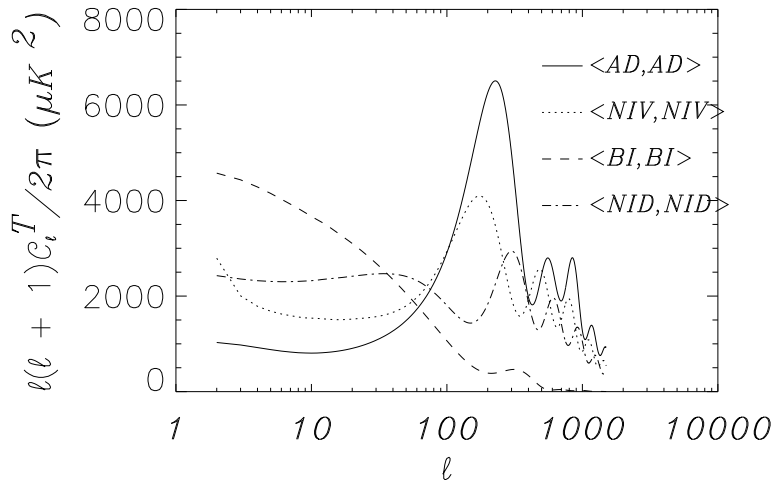,height=3.0in}}
%,height=3.5in,width=3.5in}}
%\vspace{10pt}
\caption{CMB spectra for adiabatic and isocurvature modes with cosmological
parameters $\Omega
_b=0.06,~\Omega_\Lambda=0.69,~\Omega_{cdm}=0.25,~h=0.65,~\tau_{reion} =0.1$ and $n_s=1.$}
\label{fig1}
\end{figure}

Now that we have described the five regular modes, we turn to
describing the most general primordial cosmological perturbation in a
universe with the matter content given above. We momentarily assume
Gaussianity, an assumption that we will soon be able to relax somewhat.
For a single adiabatic mode describing a Gaussian random process that
is homogeneous and isotropic, the statistical properties are
completely described by the power spectrum $P({k})$, a real valued
function of wavenumber. In the case here of five modes, whose amplitudes
we indicate as $A_a({\bf k})$, ($a=1,...,5$), the power spectrum
generalizes to a $5\times5$, positive-definite, real, symmetric,
matrix-valued function of the wavenumber $P_{ab}({k})$ where 
$\langle A_a({\bf k})~A_b(-{\bf k}')\rangle = 
P_{ab}({k})\delta^3({\bf k}-{\bf k}')$. If the fluctuations are
non-Gaussian, there are nontrivial 
higher-order correlations to consider as well,
but when observables that are quadratic in the small
perturbations are considered, $P_{ab}({k})$ suffices to
characterize their expectation values completely. 
 
\begin{table}[!t]
\centering
\begin{tabular}{|| c | c | c| c | c | c | c | c | c ||}
\hline
& MAP & MAP & MAP & MAP & PLANCK & PLANCK & PLANCK & PLANCK\cr
& T & TP & T & TP & T & TP & T & TP\cr
& adia & adia & all & all & adia & adia & all & all\cr
& only & only & modes & modes & only & only & modes & modes\cr
\hline
$\delta h/h$ &      12.37 &      7.42 &      175.84 &      20.40 &
      9.93 &      3.69 &      40.13 &        4.36\cr
\hline
$\delta \Omega _b/\Omega _b$ &      27.76 &      13.34 &      325.38 &
      28.57 &      19.37 &      7.26 &      68.85 &     
      8.61\cr
\hline
$\delta \Omega _k$ &      9.79 &      2.72 &      75.32 &      4.55
 &      4.92 &      1.83 &      20.56 &            2.18\cr
\hline
$\delta \Omega _\Lambda /\Omega _\Lambda $ &      12.92 &      5.02 &
      123.63 &      18.53 &      2.74 &      1.21 &      5.93 &
            1.49\cr
\hline
$\delta n_s/n_s$ &      7.02 &      1.62 &      89.89 &      6.53 &
     0.73 &     0.37 &      3.92 &         0.70\cr
\hline
$\tau _{reion}$ &      37.39 &      1.81 &      104.81 &      2.23 &
      8.25 &     0.41 &      35.35 &          0.56\cr
\hline
$\langle NIV,NIV\rangle $&.. &.. &      114.34 &      11.47 &.. &.. &
      43.45 &           1.14\cr
\hline
$\langle BI,BI\rangle $ &.. &.. &      573.46 &      29.71 &.. &.. &
      53.29 &            4.23\cr
\hline
$\langle NID,NID\rangle $ &.. &.. &      351.79 &      29.87 &.. &.. &
      19.18 &           2.37\cr
\hline
$\langle NIV,AD\rangle $ &.. &.. &      434.70 &      44.06 &.. &.. &
      121.59 &           4.69\cr
\hline
$\langle BI,AD\rangle $ &.. &.. &      1035.02 &      59.25 &.. &.. &
      58.75 &            8.97\cr
\hline
$\langle NID,AD\rangle $ &.. &.. &      1287.60 &      67.49 &.. &.. &
      114.39 &           5.77\cr
\hline
$\langle NIV,BI\rangle $ &.. &.. &      601.70 &      32.29 &.. &.. &
      46.91 &        3.67\cr
\hline
$\langle NIV,NID\rangle $ &.. &.. &      744.00 &      46.46 &.. &.. &
      80.01 &        2.97\cr
\hline
$\langle BI,NID\rangle $ &.. &.. &      534.32 &      39.11 &.. &.. &
      100.97 &           4.60\cr
\hline
\end{tabular}
\vskip 12pt
\caption{Percentage errors on the cosmological parameters and
the amplitudes of isocurvature auto-correlation and cross-correlation 
modes as measured by MAP and PLANCK.
Here `T' signifies that only temperature information is used while
`TP' signifies that temperature, polarization and cross-correlation 
information is used.}
\end{table}

The off-diagonal elements of $P_{ab}({k})$ represent correlations
between the various modes. In multi-field inflationary models these
generically occur. In an inflationary model with five or more fields
(or a single inflaton with the equivalent number of components), the
five fields control the amplitudes of five principal components
comprising linear combinations of the five modes. But there is no
reason why the directions of these principal components should align
with those given here for the individual modes. Therefore,
correlations generically occur \cite{langl}.

Because CMB data of the quality required to place reasonably stringent
constraints on the amplitudes of such modes (or to detect their
presence) is not yet available, it is not possible to anticipate the
broad range of possibilities for what one might discover in the new
data. But a rough estimate of the ability of new experiments (in
particular MAP \cite{mapp} and PLANCK \cite{plnck})
to constrain or detect these modes may be
obtained by assuming an underlying cosmological model with only an
adiabatic mode to produce the CMB sky (with cosmic variance, of
course) and an anticipated error model for the detector noise of these
experiments. We then determine the expected likelihood function
approximated to quadratic order about its maximum and use this
information to compute the errors on the various mode amplitudes and the
cosmological parameters when various combinations of isocurvature
modes and their correlations are admitted. A more detailed account of
this work is given in \cite{BMTb}. Some related work can be found in
\cite{related}. 

Our results are summarized in Table II. Both the MAP and PLANCK
experiments are included, with and without polarization information,
and with just the adiabatic mode and with the three isocurvature modes and
their correlations. The amplitudes of the isocurvature modes are
normalized to give the same total CMB power from $\ell=2$ through
$\ell=1500$ as the adiabatic mode. In all cases a scale-invariant
spectrum (i.e., $P(k) \sim k^{-3}$) was assumed. Tensor modes have not
been included. These would only increase the computed uncertainties.
With just the adiabatic mode
allowed, the errors on the cosmological parameters are as indicated
elsewhere in the literature. With just one extra isocurvature mode and
its correlation with the adiabatic mode added, these errors increase
modestly, in all cases by less than a factor of two.
But when three isocurvature
modes and their correlations are allowed, the uncertainties become enormous.
Errors of order unity imply a breakdown of the quadratic approximation.
The above results demonstrate that polarization information will play
a crucial role in testing adiabaticity.

\end{document}